\documentclass[12pt,preprint]{aastex}

\newcommand{\ha}{H$\alpha$}

\slugcomment{Accepted to ApJ}

\shorttitle{Thirty New Low-Mass SBs}
\shortauthors{Shkolnik et al.}

\begin{document}


\title{Thirty New Low-Mass Spectroscopic Binaries\altaffilmark{1}\\}


\author{Evgenya~L.~Shkolnik\altaffilmark{2}}
\affil{Department of Terrestrial Magnetism, Carnegie Institution of Washington, 5241 Broad Branch Road, NW, Washington, DC 20015}
\email{shkolnik@dtm.ciw.edu}

\author{Leslie Hebb}
\affil{School of Physics and Astronomy, University of St. Andrews, North Haugh St Andrews, Fife Scotland KY16 9SS, and Department of Physics and Astronomy, Vanderbilt University, Nashville, TN 37235, USA}
\email{leslie.hebb@vanderbilt.edu}

\author{Michael C. Liu\altaffilmark{3}}
\affil{Institute for Astronomy, University of Hawaii at Manoa\\ 2680 Woodlawn Drive, Honolulu, HI 96822}
\email{mliu@ifa.hawaii.edu}

\author{I. Neill Reid}
\affil{Space Telescope Science Institute, Baltimore, MD 21218}
\email{inr@stsci.edu}

\and

\author{Andrew C. Cameron}
\affil{School of Physics and Astronomy, University of St. Andrews, North Haugh St Andrews, Fife Scotland KY16 9SS} 
\email{Andrew.Cameron@st-and.ac.uk}

\altaffiltext{1}{Based on
observations collected at the W. M. Keck Observatory, the Canada-France-Hawaii Telescope and by the WASP Consortium.  The Keck Observatory is operated as a scientific partnership between the California Institute of
Technology, the University of California, and NASA, and was made possible by the generous
financial support of the W. M. Keck Foundation. The CFHT is operated by the National Research Council of Canada,
the Centre National de la Recherche Scientifique of France, and the University of Hawaii. The WASP Consortium consists of astronomers primarily from the Queen's University Belfast,
St Andrews, Keele, Leicester, The Open University, Isaac Newton Group La Palma and
Instituto de  Astrof{\'i}sica de Canarias. The SuperWASP Cameras were constructed
and operated with funds made available from Consortium Universities and the UK's Science and
Technology Facilities Council.}
\altaffiltext{2}{Carnegie Fellow}
\altaffiltext{3}{Alfred P. Sloan Research Fellow}

\begin{abstract}

As part of our search for young M dwarfs within 25 pc, we acquired high-resolution spectra of 185 low-mass stars compiled by the NStars project that have strong X-ray emission. By cross-correlating these spectra with radial velocity standard stars, we are sensitive to finding multi-lined spectroscopic binaries. 
We find a low-mass spectroscopic binary fraction of 16\% consisting of 27 SB2s, 2 SB3s and 1 SB4, increasing the number of known low-mass SBs
by 50\% and proving that strong X-ray emission is an extremely efficient way to find M-dwarf SBs.
WASP photometry of 23 of these systems revealed two low-mass EBs, bringing the count of known M dwarf EBs to 15. BD -22 5866, the SB4, is fully described in \cite{shko08} and CCDM~J04404+3127~B consists of a two mid-M stars orbiting each other every 2.048 days. 
WASP also provided rotation periods for 12 systems, and in the cases where the synchronization time scales are short, we used $P_{rot}$ to determine the true orbital parameters. For those with no $P_{rot}$, we use differential radial velocities to set upper limits on orbital periods and semi-major axes.  
More than half of our sample has near-equal-mass components ($q >$ 0.8). 
This is expected since our sample is biased towards tight orbits where saturated X-ray emission is due to tidal spin-up rather than stellar youth. 
Increasing the samples of M dwarf SBs and EBs is extremely valuable in setting constraints on current theories of stellar multiplicity and evolution scenarios for low-mass multiple systems.

\end{abstract}


\section{Introduction}\label{intro}

The multiplicity of stars is an important constraint of star formation theories as most stars form as part of a binary or higher-order multiple system (e.g.~\citealt{halb03,bate09}). 
Moreover, double- (or multi-) lined spectroscopic binaries (SBs) allow precise determination of dynamical properties including the mass ratio.
The analysis of photometric and spectroscopic 
data of eclipsing binaries (EBs) provides radii, temperatures, luminosities and masses, arguably the most important stellar parameter,
for two stars with same age and metallicity \citep{Lastennet} making them vital in calibrating stellar evolutionary models.  

M~dwarf EBs are particularly important
since the radii of active M~dwarfs are known to be 10-15\% larger than
existing models predict \citep{Mercedesa,Mercedesb,Torres09}.
This discrepancy is thought to be caused by magnetic fields on active M~dwarfs
which inhibit convection \citep{Mercedesc,Chabrier} and/or missing opacity
sources in the models \citep{berg06_d}.   Further study of the problem requires the
identification and detailed study of more M~dwarf EBs with a
range of properties (e.g.~masses, activity levels, metallicities, ages).

Though M dwarfs are ubiquitous in the Galaxy, composing 75\% of known stars \citep{boch08}, their intrinsic faintness makes them difficult and costly to observe, and thus the multiplicity of M dwarfs has been a challenge to measure. A binary fraction of 57\% has been well established for G dwarfs \citep{duqu91}, and there is clear consensus that the binary fraction of M dwarfs is significantly less than that. Published values range from 25\% \citep{lein97} to 42\% \citep{fisc92} with the largest uncertainties likely due to incompleteness corrections. In addition to supplying key constraints for star formation theories, completing the M dwarf binary census of the solar neighborhood, which to date, is (near)-complete out to only 9 pc \citep{delf04}, is yet another goal in finding the nearby low-mass SBs.

\cite{delf04} presented what may be the most complete statistical study of M dwarfs, including both spectroscopic and visual binaries.  They conclude a binary fraction of 26 $\pm$ 3\% and that for M dwarfs, as for G dwarfs, the mass ratio ($q = M_B/M_A$) distribution is a function of orbital period with most shorter period binaries having near equal component masses, while wide binaries ($P_{orb}>$50 days) have a flat \emph{q} distribution. The different distribution in the two samples may point to two distinct formation mechanisms, one for long-period and one for short-period orbits.

Here we report on 30 low-mass SBs detected from 185 X-ray-selected M dwarfs in the solar neighborhood. Six of the targets were included in the Gliese catalog of which only one was previously detected to be a SB. Prior to this work, 46 M dwarf SBs were published in the literature \citep{duqu91,delf99,fisc92} with an additional 13 low-mass EBs (\citealt{shko08} and references therein and \citealt{blak08}). Thus our work increases the known sample of SBs by 50\%. 

We also searched the WASP photometric database in which we found 2 EBs and measured rotation periods of 12 binaries.  Though rotation periods of single stars are important age indicators (e.g.~\citealt{barn07}), for short-period binaries, where tidal locking is almost certain, the rotation periods offer true orbital periods instead.

\section{Sample Selection}\label{sample}

The original science goals of the X-ray-selected sample was to search for the youngest M dwarfs within 25 pc to identify the best possible stellar targets for direct imaging searches of extrasolar planets and
circumstellar disks \citep{shko09}.
The Two Micron All Sky Survey (2MASS) is optimal for finding low-mass stars, since the SEDs of cool stars peak in the near-IR (e.g.~\citealt{hawl02}). However, the $JHK$ infrared passbands provide less distinctive spectral classification of early-~and mid-M dwarfs (i.e.~M2--M7 dwarfs have ($J-K$) colors which span only 0.2 mag; \citealt{reid07a}) impeding the photometric distance determination. Thus, in order to fully characterize a volume-limited sample of young M dwarfs, a proper motion requirement of $\mu >$0.18\arcsec\ yr$^{-1}$ was implemented, equivalent to a tangential velocity of 21 km~s$^{-1}$ at 25 pc.

We drew $\approx$800 targets from the NStars 20-pc census \citep{reid03,reid04} constructed from the 2MASS catalogs (\citealt{skru06}) along with the \cite{lepi02} and \cite{lepi05} proper motion catalogs, with an additional $\sim$300 newly-catalogued M dwarfs that exhibit significant proper motion between the POSSI and 2MASS surveys (i.e.~the Moving-M sample; \citealt{reid07b}). Distances were available either from parallaxes or spectrophotometric relations and are limited to 25 pc from the Sun, good to $\lesssim$15\% assuming all the stars were single and on the main-sequence (e.g., \citealt{reid02c,cruz03}). 
Further details of the sample selection can be found in \cite{shko09}.

\begin{figure}
\epsscale{1.0}
\plotone{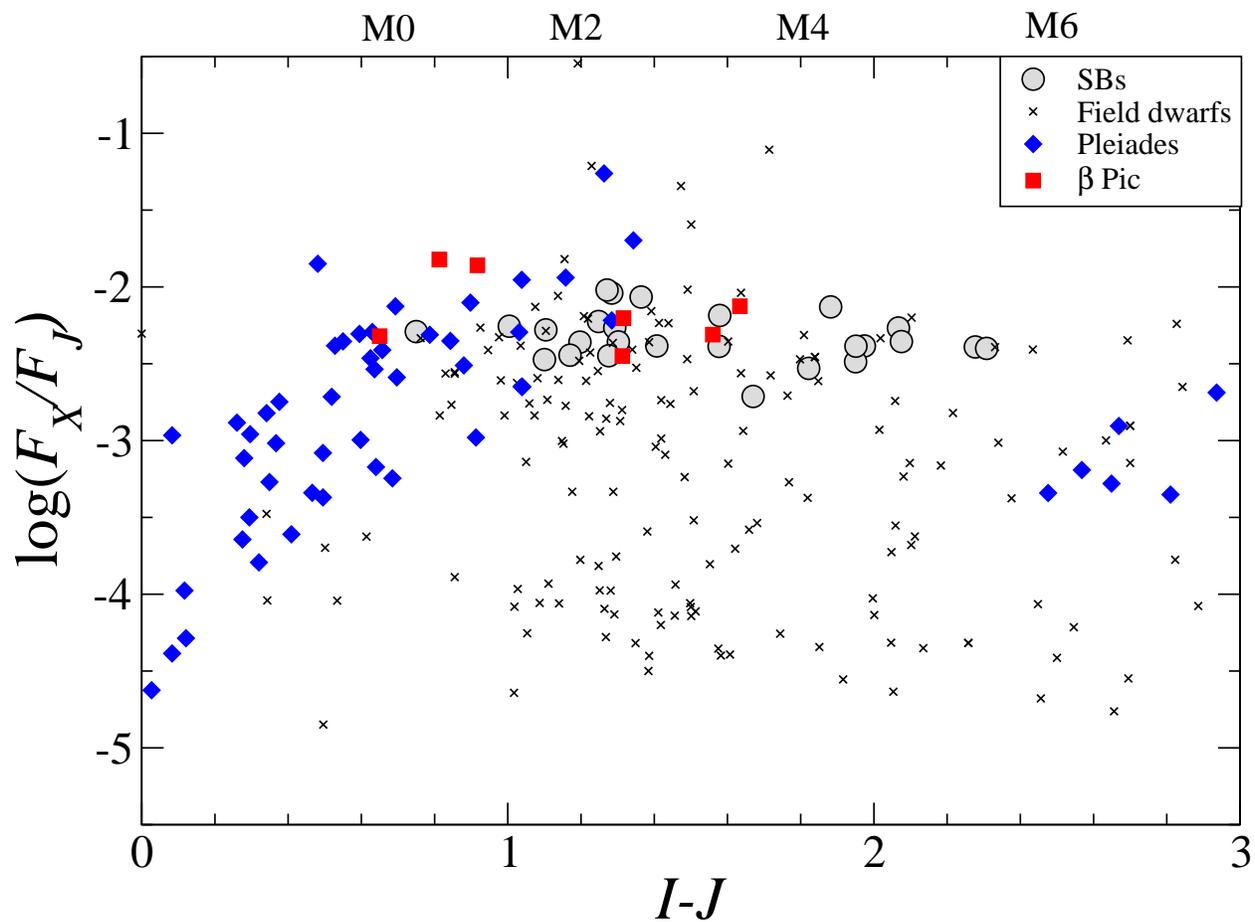}
\caption{The fractional X-ray luminosity as a function of $I-J$ for our sample of SBs compared with seven members of the $\beta$ Pic young moving group at 12 Myr \citep{torr06}, Pleiades members at 120 Myr (\citealt{mice98}), field stars (\citealt{huns99}).
\label{ij_fxfj}}
\end{figure}

In order to characterize the youngest members of this sample, we acquired high-resolution spectra of the 185 cool dwarfs with bright X-ray luminosities. In Figure~\ref{ij_fxfj} we plot the fractional X-ray flux, $F_X/F_J$ as a function of $I-J$, where $F_X$ is the empirically calibrated \emph{ROSAT} X-ray flux \citep{voge99} using the count-rate conversion equation of \cite{schm95}, and $F_J$ is the 2MASS $J$-band flux. Target stars were chosen to have high X-ray emission ($log(F_X/F_J) > -2.5$) near the saturation limit 
\citep{riaz06} and comparable to or greater than the fractional luminosities of Pleiades members (120 Myr, \citealt{mice98}) and $\beta$~Pic members (12 Myr, \citealt{torr06}). 
We measured the radial velocities (RVs) needed to determine galactic space motion and young moving group membership (Shkolnik et al., in prep.).  This process is also sensitive to finding SBs at any age, particularly those in short-period orbits where the tidal-locking forces rapid rotation producing high chromospheric and X-ray emission.

\section{The spectra}\label{spectra}

We acquired spectra with the High Resolution \'Echelle Spectrometer (HIRES; Vogt et al.~1994) on the Keck I 10-m telescope and the \'Echelle SpectroPolarimetric Device for
the Observation of Stars (ESPaDOnS; Donati et al.~2006) on the Canada-France-Hawaii 3.6-m telescope, both located on the summit of Mauna Kea.

We used the 0.861$\arcsec$ slit with HIRES to give a spectral resolution of $\lambda$/$\Delta\lambda$$\approx$58,000. The detector consists of a mosaic of three 2048 x 4096 15-$\micron$ pixel CCDs, corresponding to a blue, green and red chip spanning 4900 -- 9200 \AA. To maximize the throughput near the peak of a M dwarf spectral energy distribution, we used the GG475 filter with the red cross-disperser. The data product of each exposure is a multiple-extension FITS file from which we reduced and extracted the data from each chip separately.  

ESPaDOnS is fiber fed from the Cassegrain to Coud\'e focus where the fiber image is projected onto a Bowen-Walraven
slicer at the spectrograph entrance. With a 2048$\times$4608-pixel CCD detector, ESPaDOnS'
`star+sky' mode records the full spectrum over 40 grating orders covering 3700 to 10400 \AA\/ at a spectral
resolution of $\lambda$/$\Delta\lambda$$\approx$68,000. The data were reduced using {\it Libre Esprit}, a fully automated reduction package
provided for the instrument and described in detail by Donati et al.~(1997, 2007).

\begin{figure}
\epsscale{0.7}
\plotone{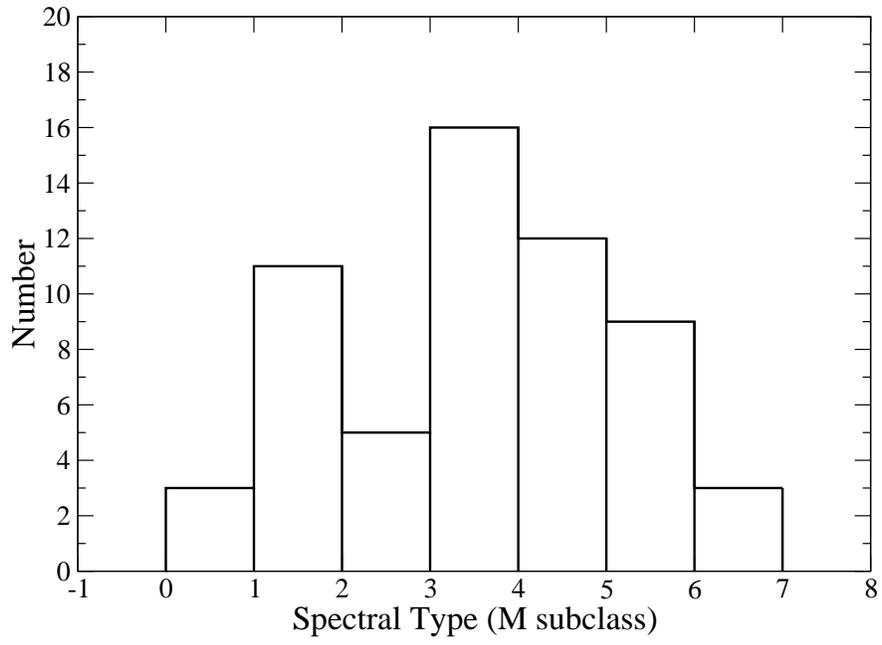}
\caption{Histogram of the spectral types of the component stars. 
\label{SpT_hist}}
\end{figure}

All final spectra were of moderate S/N reaching $\approx$ 50 per pixel at 7000 \AA. Each night, spectra were
also taken of an A0V standard star for telluric line correction and an early-, mid- and late-M RV standard. 
To search for multi-lined binaries, we cross-correlated each of 7 orders between 7000 and 9000 \AA\/ of each stellar spectrum (from both instruments) with a RV standard of similar spectral type using IRAF's\footnote{IRAF (Image Reduction and Analysis Facility) is distributed by
the National Optical Astronomy Observatories, which is operated by the Association of Universities for Research in Astronomy, Inc.~(AURA) under cooperative agreement with the National Science Foundation.} {\it fxcor}
routine (Fitzpatrick 1993).\footnote{See Table~3 of \cite{shko09} for the list of RV standards used.}
We measure the RVs of each component from the gaussian peaks fitted to the cross-correlation function  (CCF), taking the average of all orders, with a RMS typically less than 1 km~s$^{-1}$ for both instruments. 
The resolved CCF peaks allowed us to estimate the spectral types, and thus component masses, of the individual stars assuming a flux-weighted relation between component and integrated spectral types (\citealt{daem07}): $\mathrm{SpT_{int}} = (f_\mathrm{A} \mathrm{SpT_\mathrm{A}} + f_\mathrm{B} \mathrm{SpT_\mathrm{B}})/(f_\mathrm{A}+f_\mathrm{B})$. Evidence for this linear relationship is presented by \cite{cruz02}, which shows both the TiO and VO band depths vary linearly with SpT for stars ranging from M0 to M7. 
$\mathrm{SpT_{int}}$ was measured from the ratio of the band indices of TiO $\lambda$7140 to TiO~$\lambda$8465 defined by \cite{sles06} and calibrated with 136 M dwarfs of known SpT \citep{shko09}.  $f_\mathrm{A}$ and $f_\mathrm{B}$ were derived from the integrated ``flux'' of the gaussian fits to the cross-correlation peaks. An additional restriction to the spectral types of the two components is provided by $\Delta R$, where $\Delta R = M_R(\mathrm{SpT_\mathrm{B}})- M_R(\mathrm{SpT_\mathrm{A}})$. These two equations together produce a unique solution for the component spectral types, which are good to 0.5 subclasses or better.
The integrated properties of the binary systems, including the systemic velocity $\gamma$ and the corrected photometric distance, are listed in Table~\ref{integrated_data} with a histogram of the component SpTs in Figure~\ref{SpT_hist}. We can estimate the mass of the components from these SpTs using the mass relationship from \cite{reid05}, which is based on the 8-pc sample of M dwarfs originally presented in \cite{reid97}.  The error on the masses ranges from 10\% for M0 to 40\% for M6, mostly due to the error in spectral typing the two components.

\section{WASP Photometry}\label{photometry}

We cross-correlated our sample of M~dwarf SBs with the UK Wide-Angle Search for Planets (WASP)
database \citep{swasp} and identified 23 observed systems, of which 12 have clear rotational modulation, including 2 new EBs.

WASP is a wide-field photometric variability survey of bright
stars designed to detect significant numbers of transiting gas giant planets.
The survey operates two robotic telescopes, one in each hemisphere, which 
repeatedly observe nearly the entire visible sky every clear night.  
The resulting data product of the ongoing survey consists of 
high cadence ($\sim 8$~minute sampling), precise ($\sim 1$\%
r.m.s.), single-band light curves for millions of bright ($V=9-13$) stars with spectral types that
cover the Hertzsprung-Russell diagram.  Each stellar light curve typically has several thousand
photometric data points obtained over several years with typical photometric precision of 0.007 -- 7 mag per observing season for the above magnitude range.  
The stars that are observed exist in all regions of the sky 
except the Galactic plane due to crowding and extinction and the equatorial poles, 
which have yet to be surveyed.  The WASP survey 
is the most successful transiting planet survey to date and has discovered over 20 new 
transiting gas giant planets \citep[e.g.][]{wasp1,wasp5,wasp15,wasp12,wasp18}. However, for our SB targets, we are able to detect rotational variability with an amplitude > 0.01 mags.

\subsection{Rotation Periods}

Asymmetrically distributed starspots on the photosphere of one or more
components of the binary will cause periodic brightness variations as the spots 
rotate in and out of view with the star.  The stellar rotation period (or a harmonic)
can be identified through the detection of a periodic sinusoidal variability signal.  
Starspots typically evolve on timescales of weeks to months which leads to
changes in the amplitude and/or phase of the variability, thus we independently 
searched each season of photometric WASP data to maximize our chances of 
detecting a periodic signal.

To measure the level of periodic variability in each target, we determined the improvement in $\chi^2$ over
a flat, non-variable model ($\delta\chi^2)$ when a sine wave of the form  
$y = a_0 + a_1 sin(\omega t + a_3)$ was fit to each season of the
WASP data phase-folded at a set of trial periods, $P_{rot}=2\pi/\omega$.
Periods between 0.2--30~days were searched.
The statistic, $\delta\chi^2/\chi^2_{best}$ \citep{zechkurst}, where
$\chi^2_{best}$ is the $\chi^2$ with respect to the best fitting sinusoidal
model, was used to distinguish genuine variability in each light curve.  
Our sensitivity to the period detection is different for each object, but in
general, we can detect a sinusoidal signal with an amplitude of $> 5$~mmag and a
period, $P_{rot} < 30$~days. 

\begin{figure}
\epsscale{1.0}
\includegraphics[angle=90,width=6in]{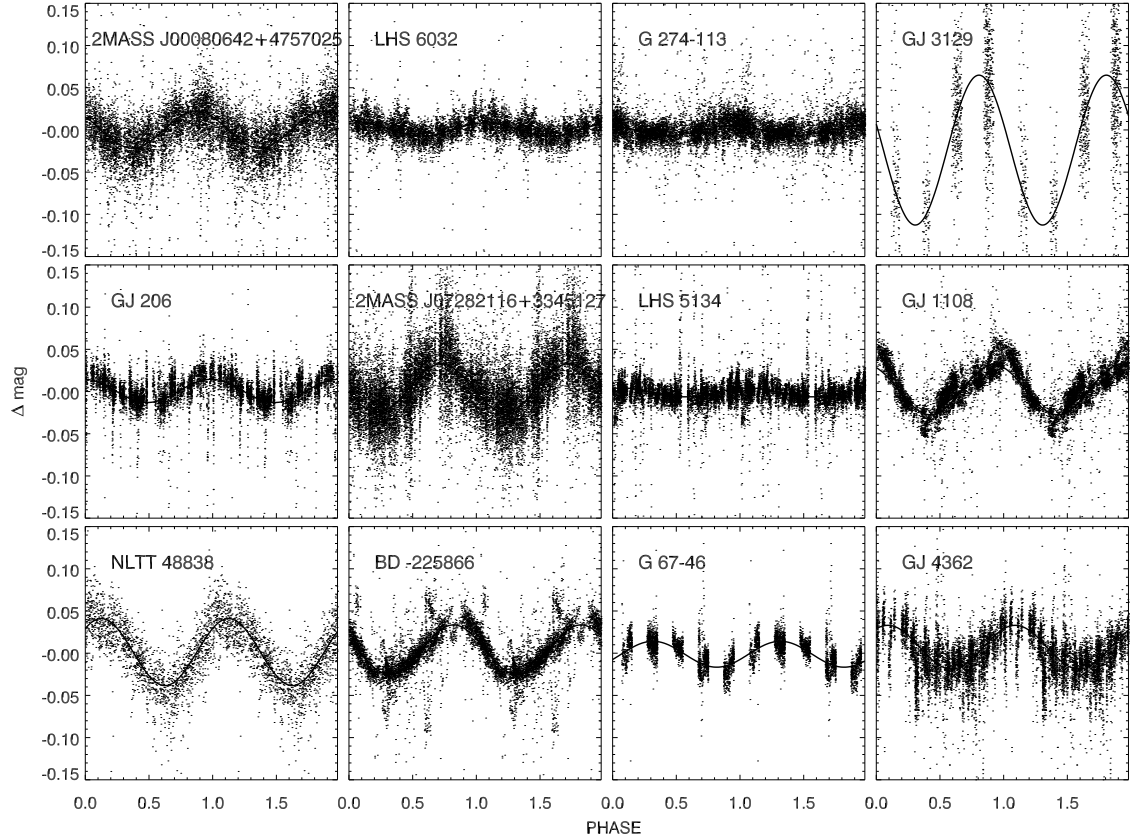}
\caption{
Phase-folded WASP light curves of the 12 M~dwarf SBs for which rotation
periods are detected.  The rotation period listed in Table~\ref{targets_with_Prot} is used
to convert HJDs to phase values.  The best fitting model sine curve for each target is
over plotted on the phase-folded data.   Note: if more than one season of WASP
data exists, the one with the strongest signal is shown. 
\label{swasplc}}
\end{figure}

Twelve of the 23 M~dwarf SBs with WASP data\footnote{Seven of the SBs have not yet been sufficiently observed by WASP or are not accessible to the telescopes.} show periodic sinusoidal variability
in at least one season of WASP data.  
These are listed in Table~\ref{targets_with_Prot} and their phase-folded light curves
are shown in Figure~\ref{swasplc}.  We note that four of the systems 
have multiple seasons of data in which the same or a similar period is detected 
which provides additional confirmation of the period detection.
No variability is seen in the 
remaining 11 targets for which we have time-series data (Table~\ref{targets_without_Prot}).
In both tables, we first list the target and the number
of data points available in each season as well as the HJD of the starting
and ending dates for the light curve.  We also list the value of the $\delta\chi^2/\chi^2_{best}$
statistic for the highest peak in the periodogram.  All the positive detections have
$\delta\chi^2/\chi^2_{best} \ge 0.29$ and the non-detections typically have $\delta\chi^2/\chi^2_{best} < 0.15$.  
For the positive detections, we also list the rotation period and the amplitude of the variability.

\subsection{Low-mass eclipsing binaries}

In addition to discovering transiting extrasolar planets, the time-series photometry obtained through
the WASP survey is an excellent data set for identifying eclipsing binary stars. Therefore, we searched
the WASP light curves of the SB sample for periodic eclipses. We employed the WASP implementation of
the box-least squares (BLS) algorithm \citep{kovacs,hunter} which is designed to detect square-shaped
dips in brightness in an otherwise flat light curve. The BLS algorithm is efficient at detecting both
transiting planets \citep{tingley} and EBs \citep{hartman}. We searched periods between 0.5-10~days
and detected two eclipsing binaries. BD~-22~5866 is a late-type quadruple system composed of two
binary pairs (K7+K7 and M1+M2) in which the more massive binary is eclipsing with a 2.21~day period. This unique system is described in \citet{shko08}. 

\begin{figure}
\epsscale{1.0}
\includegraphics[angle=90,width=6in]{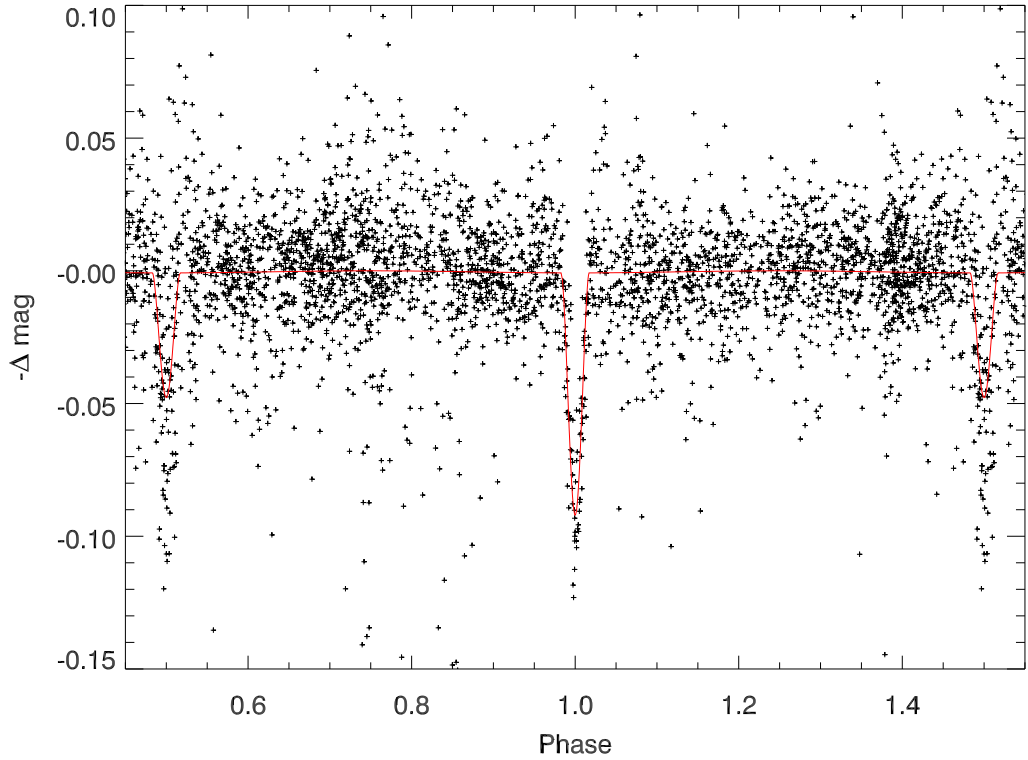}
\caption{WASP light curve of CCDM~J04404+3127B phase-folded with the ephemeris 
$(2454129.2969 \pm 0.0007) + (2.048135 \pm 0.000003) E$.  Over-plotted (red line) is the best 
fitting EBOP model.  There is no detectable rotational variability
in the light curve.  We also note that the eclipsing binary, CCDM~J04404+3127B,
is blended with CCDM~J04404+3127A in the WASP photometry causing the eclipse
depths to appear shallower than they should. 
\label{eb}}
\end{figure}

The second EB we detected
in the M~dwarf SB sample is CCDM~J04404+3127~B. 
The object is part of a triple system composed of the EB with a proper motion companion $\sim 15^{\prime\prime}$ away (CCDM J04404+3127~A).
The photometry shows an eclipsing light curve for CCDM~J04404+3127 (Figure~\ref{eb}), but since the A and B components of
the multiple system are blended in the large photometric aperture ($\sim 48^{\prime\prime}$), 
the WASP data alone cannot determine which component is the unresolved EB.\footnote{\citet{hartman} also identify this object
(HAT-216-0003316) as a probable EB based on HAT transiting planet survey data, but the object is
blended with CCDM~J04404+3127~A in their data as well.}
However, given that CCDM~J04404+3127~A does not have a ROSAT detection and the single spectrum of CCDM~J04404+3127~B presented here confirms it be a spectroscopic binary, it must also be the EB.  

The target was observed by the SuperWASP camera on La~Palma, Canary Islands in the 2004 and 2006 
observing seasons.   The existing photometric data consists of 949 measurements taken 
between 1~August and 29~September 2004 and 2478 measurements taken 
from 29~September 2006 to 15~February 2007.  After detecting the eclipsing binary 
using the BLS algorithm described above, we fit a more realistic eclipsing light curve model
to the combined light curve to derive an accurate ephemeris for the system.  We used the JKT
Eclipsing Binary Orbit Program (EBOP; \citealt{popperetzel,southworth} to derive an ephemeris
of:\footnote{The ephemeris reported here is consistent 
with the period determined by \citet{hartman}.} $$ {\rm Min(HJD)} = (2454129.2969 \pm 0.0007) + (2.048135 \pm 0.000003) E$$
\noindent Due to the contamination by light from CCDM J04404+3127~A, we cannot confidently report 
any other parameters for the system until more data is collected. However from the spectrum, the EB consists of a M4.2 and a M5.0 dwarf.

\section{Binary Properties}\label{results}

\subsection{Tidal Synchronization and Maximal Orbits}

With a single spectroscopic observation, the difference between the radial velocity of each component ($\Delta$RV) provides an upper limit to the orbital period and semi-major axis using the stellar masses determined from the integrated fluxes (i.e.~the area) of the cross-correlation peaks. However, the added photometry of WASP can provide a true orbital period and semi-major axis if the stellar system is expected to be tidally synchronized, i.e. $P_{orb} = P_{rot}$.  


The tidal synchronization time scale $t_{sync}$ of close-in binary systems is fairly short, e.g. less than 200 Myr for two 0.5-M$_{\odot}$ stars with $P_{orb}$=5 days as calculated using Equation 6.1 of \cite{zahn77}. \cite{drak98} show that Zahn's values serve as an upper limit to the time needed to synchronize the upper envelopes of the stars and thus the time that it would take for $P_{rot}$ to equal $P_{orb}$ may be $>$ 3 times shorter than $t_{sync}$. Such time scales are comparable to the lithium depletion time scale, which ranges from 15 Myr for a M0 to 90 Myr for a M6 dwarf \citep{chab96}.  We do not detect any lithium in these stars, nor any other spectroscopic indication that the stars are young (e.g.~low surface gravity and strong \ha\/ emission), and thus the stars are almost certainly older than $> 0.2$~Gyr.  

Of the 12 binaries with measured rotation periods, 2 have calculated $t_{sync}$ (using $a_{max}$) of less than 200 Myr. For these systems, we assume that the  binaries are synchronized such that the observed photometric period is equal to both the rotation \emph{and} orbital periods. 
These periods and corresponding semi-major axes are listed in Table~\ref{binary_data}. For those stars where there is no photometric data or the photometric period is long enough such that if it were the orbital period $t_{sync} \gg 10$ Gyr, we list the maximal orbital parameters, $P_{max}$ and $a_{max}$. 

With only one $\Delta$RV measurement, we cannot assume tidal synchronousity in those systems with short rotational periods but with long $t_{sync}$.  The rapid rotation of one or both of the components may be due to a young system in a wider orbit. Although we can set some lower limits on the ages of the stars using surface gravity indicies and the lack of lithium absorption, these limits are generally short compared to the spin-down rate of M dwarfs. \cite{delf98} reported that early M's spin down to 2~km~s$^{-1}$ or less (or $P_{rot}$ = 10 -- 15 days) by 500 Myr while late Ms take $\sim$1 Gyr to spin down to 2 km~s$^{-1}$. These time scales are comparable to the statistical investigations of \cite{west08}.  And, given that the sample was selected for strong X-ray luminosity, we cannot rule out that our SB sample maybe indeed contain  young M dwarfs in wide binary systems. However, it remains probable that all the stars in this sample will eventually be shown to be close-in, rapidly orbiting and rotating SBs.

\subsection{X-ray saturation in short-period systems}

\begin{figure}
\epsscale{1.0}
\plottwo{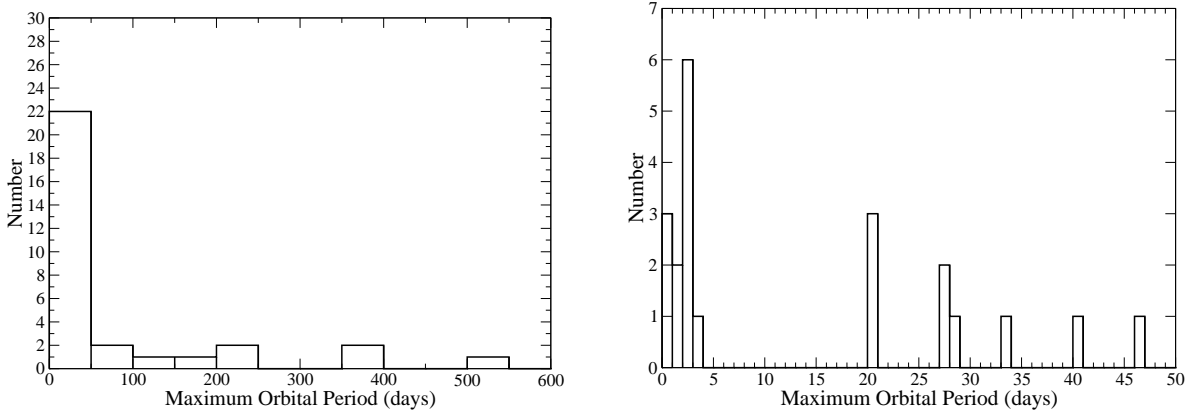}{f5b.eps}
\caption{\emph{Left:} Histogram of orbital periods consisting of $P_{max}$ and $P_{orb}$ if available. \emph{Right:} An expanded histogram of the tightest orbits. 
\label{Porb_hist}}
\end{figure}

At least 12 of the 30 systems have orbital periods less than 4.5 days (Figure~\ref{Porb_hist}) clearly displaying our
bias towards tidally spun-up stars. Data compiled by \cite{riaz06} of a large set of M dwarfs suggest that
the coronal activity remains at a high level independent of SpT and age
implying that rotation, not age, dominates the magnetic activity in low-mass stars.

Recently, \cite{rein09} measured that in M dwarfs the surface magnetic flux also saturates, similar to the activity implying that a star with saturated levels of X-ray emission does not necessarily have a large filling factor (i.e. the whole star need not be covered with active regions). This is supported by the clear rotation periods we detect.  However, there are three targets, all of which are mid-Ms, (2MASS~J0808+4347, HAT~199-13890, and LSPM J2114+1254) for which there is no modulated photometric signal and yet the maximum orbital periods are quite short ($<$ 3 days).  In these cases, we suspect that the filling factor is indeed high, such that the symmetric spots do not produce any photometric variability. This is supported by spectropolarimetric observation by \cite{mori08} who report that a sample of 5 mid-M stars mainly host axisymmetric large-scale poloidal fields.

\subsection{Mass Ratios}

\begin{figure}
\epsscale{0.7}
\plotone{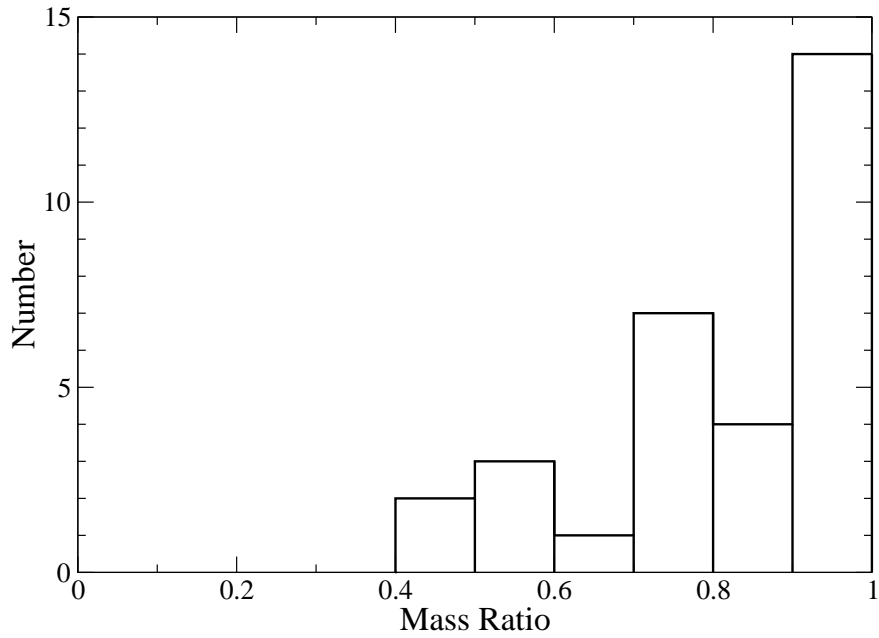}
\caption{Histogram of SB mass ratios.
\label{q_hist}}
\end{figure}

Past studies of SBs have somewhat crudely shown that the mass ratio $q$, is inversely correlated with the mass of the primary star. For solar-type stars, $q$ peaks at about 0.2 \citep{duqu91}, while \cite{halb03} found a bimodal distribution for A to K type stars: $q$ has a broad distribution between 0.2 and 0.7 for long orbital periods, and a strong peak with near equal-mass systems, primarily in shorter orbital period ($<$ 100 days).  For M dwarfs, \cite{fisc92} measured a flat distribution for $q$ over all periods in a sample of visual and spectroscopic binaries. Although our range for $q$ matches theirs (0.4 -- 1; Figure~\ref{q_hist}), more than half of our sample has $q >$ 0.8, which is not the case in the \cite{fisc92} sample.
This is expected since our sample is biased towards tight orbits where the primary selection criterion of saturated X-ray emission is due to tidal spin-up. In addition, these binaries with large $q$ will synchronize much more quickly as $t_{sync} \propto q^{-2}$.

\section{Conclusions}\label{summary}

Of our sample of 185 X-ray bright M dwarfs, we find a low-mass, multi-lined spectroscopic binary fraction of 16\%.
These 30 SBs are composed of 27 SB2s, 2 SB3s and 1 SB4, increasing the number of known
low-mass SBs by 50\% and proving that strong X- ray emission is an extremely efficient way to find
M-dwarf SBs.

To search for single-lined SBs (SB1), we observed two epochs (separated typically by 2--3 months) of 65 of the 185 targets, none of which showed a
significant RV variation between visits to the level of 1 km~s$^{-1}$. This implies that the single-lined binary fraction of stars with orbital periods of less than about 1.5 years in our
sample is very low, less than 1.5\%, and that M dwarf binaries with low mass ratios ($q \ll 1$) are
rare. It is possible that up to 4\%\footnote{This 4\% limit is based on the time a close-in low-mass binary with an orbital period of 5 days
would spend near conjunction such that the RVs of the two components would not produce resolved peaks
in the CCF.} of the stars with a single observation are indeed double-lined SBs if the systems were in
conjunction at the times of the observation. Combining this with the $\leq$1.5\% chance of observing
an SB1, there are at most a handful of undiscovered SBs in the original 185
ROSAT-selected targets, setting an upper limit of 19\% to the SB fraction (with $P_{orb} \lesssim 1.5$ years) of our sample.

WASP photometry of 23 of these systems revealed two low-mass EBs, bringing the count of known M dwarf EBs to 15. The WASP data also provided rotation periods for 12 systems, and in the cases where the synchronization time scales are short, orbital periods and semi-major axes. 

This X-ray bright sample of 30 SBs is strongly biased towards high-$q$, tidally-synchronized binaries. In addition to being in short-period orbits, they are also relatively bright, making them excellent targets for a spectroscopic monitoring program to measure the component velocities necessary to determine the Keplerian orbital parameters for more precise mass ratios, and in the case of the eclipsing systems, the individual masses needed to test evolutionary models.

\acknowledgements

We thank the anonymous referee for her/his insightful comments on the original manuscript. E.S also thanks the CFHT and Keck staff for their care in setting up the instruments
and support in the control room. This material is based upon work supported by the Carnegie Institution of Washington and the National Aeronautics and Space Administration through the NASA
Astrobiology Institute and the NASA/GALEX grant program under Cooperative Agreement Nos. NNA04CC08A and NNX07AJ43G issued through the Office of
Space Science. M.C.L. acknowledges support from the Alfred P. Sloan Research Fellowship.

\clearpage
\bibliography{refs_v2}{}
\bibliographystyle{apj}

\clearpage 

\begin{deluxetable}{llccccrlllcccc}
\tabletypesize{\scriptsize}
\rotate
\tablecaption{M dwarf Spectroscopic Binaries\label{integrated_data}}
\tablewidth{0pt}
\tablehead{
\colhead{Name} & \colhead{RA \& DEC} & \colhead{SpT$_{int}$}  &  \colhead{$I$} & \colhead{$J$} & \colhead{log$(F_X/F_J)$\tablenotemark{a}} & \colhead{$\gamma$} & \colhead{HJD} & \colhead{Dist.\tablenotemark{b}} & \colhead{Binarity\tablenotemark{c}}\\
\colhead{} & \colhead{J2000} & \colhead{M--($\pm 0.5$)} & \colhead{$_{USNO}$} & \colhead{$_{2MASS}$} & \colhead{} & \colhead{km~s$^{-1}$} & \colhead{--2450000}  & \colhead{pc} & \colhead{}

}
\startdata
							
2MASS J00080642+4757025	&	00	08	06.4	+47	57	02.0	&	3.6	&	10.1	&	8.523	&	-2.387	&	-29.64	$\pm$	0.44	&	3961.10494	&	16.2	$\pm$	1.9	&	SB2	\\
LHS 6032                	&	01	45	18.2	+46	32	07.8	&	1.7	&	9.94	&	8.058	&	-2.130	&	17.85	$\pm$	0.69	&	4288.12715	&	26.7	$\pm$	2.7	&	SB2, VB (NE)	\\
G 274-113               	&	01	53	11.3	-21	05	43.0	&	1.3	&	10.04	&	8.066	&	-2.384	&	14.48	$\pm$	0.51	&	4378.96250	&	28.6	$\pm$	2.0	&	SB2	\\
NLTT 6638               	&	01	59	12.6	+03	31	11.3	&	2.5	&	--	&	7.998	&	-2.413	&	-9.48	$\pm$	0.74	&	4288.13890	&	11.2	$\pm$	2.7	&	SB2, VB (NE)	\\
GJ 3129                 	&	02	02	44.2	+13	34	33.0	&	4.7	&	10.9	&	9.652	&	-2.225	&	-12.94	$\pm$	0.60	&	4378.98460	&	22.7	$\pm$	2.2	&	SB2	\\
GJ 3236                 	&	03	37	14.1	+69	10	49.8	&	4.1	&	10.91	&	9.806	&	-2.279	&	16.95	$\pm$	3.55	&	4155.81744	&	22.0	$\pm$	2.7	&	SB2	\\
NLTT 11415              	&	03	37	33.4	+17	51	14.5	&	1.9	&	9.85	&	9.1	&	-2.291	&	40.22	$\pm$	0.51	&	4155.82919	&	31.3	$\pm$	2.7	&	SB2, VB of GJ3240B	\\
GJ 3240 B               	&	03	37	33.9	+17	51	00.4	&	4.3	&	10.19	&	9.186	&	-2.256	&	38.25	$\pm$	3.57	&	3725.98033	&	21.9	$\pm$	2.7	&	E?SB2, VB	\\
2MASS J04244260-0647313 	&	04	24	42.6	-06	47	31.0	&	5.2	&	10.93	&	9.566	&	-2.066	&	13.05	$\pm$	0.77	&	3726.00509	&	17.9	$\pm$	1.6	&	SB3	\\
CCDM J04404+3127B       	&	04	40	23.0	+31	26	46.2	&	4.5	&	11.22	&	10.023	&	-2.359	&	41.73	$\pm$	1.19	&	3726.03777	&	30.5	$\pm$	2.7	&	ESB2, VB	\\
GJ 206	&	05	32	14.7	+09	49	15.0	&	4.1	&	9.7	&	7.423	&	-2.391	&	20.66	$\pm$	0.63	&	3726.08132	&	12.8	$\pm$	0.6\tablenotemark{d}	&	SB2	\\
GJ 3362	&	05	40	16.1	+12	39	00.8	&	1.4	&	10.38	&	8.072	&	-2.400	&	95.12	$\pm$	1.68	&	3726.08645	&	24.2	$\pm$	2.7	&	SB2	\\
G 108-4                 	&	06	29	50.2	-02	47	45.0	&	5.7	&	10.76	&	9.468	&	-2.266	&	83.59	$\pm$	0.38	&	3726.11372	&	37.6	$\pm$	2.2	&	SB2	\\
2MASS J07282116+3345127	&	07	28	21.2	+33	45	12.0	&	3.9	&	10.45	&	9.28	&	-2.444	&	11.34	$\pm$	1.04	&	3866.72743	&	32.7	$\pm$	3.7	&	SB2	\\
LHS 5134                	&	08	08	13.6	+21	06	09.0	&	2.6	&	--	&	7.336	&	-2.327	&	82.87	$\pm$	0.75	&	3726.12839	&	17.1	$\pm$	0.8\tablenotemark{d}	&	SB2, VB	\\
2MASS J08082487+4347557	&	08	08	24.9	+43	47	55.0	&	5.5	&	12.56	&	10.493	&	-2.267	&	72.95	$\pm$	0.57	&	3726.12411	&	22.4	$\pm$	2.6	&	SB2	\\
GJ 1108 B               	&	08	08	55.4	+32	49	04.7	&	3.0	&	--	&	7.999	&	-2.053	&	12.12	$\pm$	0.70	&	3866.77237	&	20.7	$\pm$	1.5\tablenotemark{d}	&	SB2, VB	\\
GJ 3547                   	&	09	19	22.9	+62	03	16.8	&	0.0	&	9.99	&	8.168	&	-2.530	&	68.52	$\pm$	1.21	&	3866.82837	&	32.5	$\pm$	2.8\tablenotemark{d}	&	SB2	\\
GJ 3630                   	&	10	52	03.3	+00	32	38.3	&	4.8	&	10.71	&	9.426	&	-2.040	&	25.88	$\pm$	1.35	&	3726.14401	&	19.0	$\pm$	2.7	&	SB3	\\
2MASS J12065663+7007514 	&	12	06	56.6	+70	07	51.4	&	3.9	&	9.1	&	9.251	&	-2.296	&	-21.81	$\pm$	0.53	&	4455.04585	&	16.8	$\pm$	1.9	&	SB2, VB (W)	\\
2MASS J14204953+6049348	&	14	20	49.5	+60	49	34.0	&	3.5	&	11.16	&	10.06	&	-2.473	&	-23.15	$\pm$	0.85	&	3960.75723	&	44.6	$\pm$	5.1	&	SB2	\\
GJ 616.2 	&	16	17	05.4	+55	16	09.0	&	1.2	&	8.55	&	6.6	&	-2.487	&	-28.25	$\pm$	0.92	&	3960.79603	&	20.7	$\pm$	0.5\tablenotemark{d}	&	SB2, VB	\\
HAT 199-13890           	&	19	31	12.6	+36	07	30.0	&	5.1	&	10.88	&	9.609	&	-2.020	&	-22.31	$\pm$	1.24	&	3867.04116	&	21.7	$\pm$	1.8	&	SB2, VB	\\
NLTT 48838              	&	20	10	34.5	+06	32	14.1	&	3.6	&	9.6	&	8.021	&	-2.186	&	-52.15	$\pm$	1.25	&	3960.87589	&	15.3	$\pm$	2.7	&	SB2	\\
2MASS J21021569-3129118	&	21	02	15.6	-31	29	11.0	&	4.3	&	11.26	&	9.853	&	-2.384	&	-6.98	$\pm$	0.42	&	3867.12034	&	16.9	$\pm$	2.2	&	SB2	\\
LSPM J2114+1254         	&	21	14	49.1	+12	54	00.2	&	5.5	&	11.21	&	9.908	&	-2.358	&	-58.58	$\pm$	0.50	&	3867.11455	&	27.8	$\pm$	2.7	&	SB2	\\
BD -225866	&	22	14	38.4	-21	41	53.0	&	0.0	&	9.21	&	7.54	&	-2.713	&	-14.07	$\pm$	2.43	&	3867.12294	&	40.1	$\pm$	3.2	&	ESB4 	\\
G 67-46                 	&	23	06	23.8	+12	36	26.7	&	0.7	&	10.45	&	8.375	&	-2.356	&	-3.53	$\pm$	0.96	&	3960.96632	&	34.3	$\pm$	2.7	&	SB2(3?), VB	\\
GJ 4359                 	&	23	43	59.5	+64	44	28.9	&	0.8	&	10.1	&	8.149	&	-2.386	&	14.56	$\pm$	1.08	&	3960.98538	&	28.9	$\pm$	1.7\tablenotemark{d}	&	SB2	\\
GJ 4362                 	&	23	48	36.0	-27	39	38.0	&	2.1	&	9.86	&	8.584	&	-2.448	&	25.30	$\pm$	0.55	&	3961.08236	&	24.6	$\pm$	1.9	&	SB2	\\

\enddata
\tablenotetext{a}{J band fluxes calculated here use the full 2MASS bandwidth of 0.29 $\micron$.}
\tablenotetext{b}{Photometric distances from \cite{reid02b,reid07b} were corrected for the over-luminosity of the binary system.}
\tablenotetext{c}{Those targets with directions in parentheses were resolved as visual binaries (VB) at the telescope. Those without directions, are listed as VBs in the Washington Visual Double Star Catalog \citep{worl97}.
Eclisping SBs are designated as ``ESB?''.}
\tablenotetext{d}{Distances using trigonometric parallaxes from the Hipparcos \& Tycho Catalogues (\citealt{perr97}).}

\end{deluxetable}


\begin{deluxetable}{lcclccrrrlcccc}
\tabletypesize{\scriptsize}
\tablecaption{RVs \& Orbital Parameters\label{binary_data}}
\tablewidth{0pt}
\tablehead{
\colhead{Name} & \colhead{SpT}  &  \colhead{Mass\tablenotemark{a}} & \colhead{$q$} & \colhead{RV} & \colhead{$P_{rot}$\tablenotemark{b}} & \colhead{$P_{orb}$} & \colhead{$a$} & \colhead{$t_{sync}$} \\
\colhead{} & \colhead{M--($\pm 0.5$)} & \colhead{$M_\odot$} & \colhead{} & \colhead{km s$^{-1}$} & \colhead{days} & \colhead{days} & \colhead{AU}  & \colhead{Gyr} 

}
\startdata
							
2MASS J00080642+4757025	A	&	3.6	&	0.2	&	1.00	&	17.01	$\pm$	0.33	&	4.38	&	4.4\tablenotemark{c}	&	0.04	&	0.151	\\
2MASS J00080642+4757025	B	&	3.5	&	0.2	&		&	-76.30	$\pm$	0.29	&		&		&		&		\\\\
LHS 6032                	A	&	1.9	&	0.44	&	1.00	&	36.56	$\pm$	0.53	&	4.05	&	$<$157.9	&	$<$0.56	&	$\gg$10	\\
LHS 6032                	B	&	1.6	&	0.44	&		&	-0.86	$\pm$	0.45	&		&		&		&		\\\\
G 274-113               	A	&	1.1	&	0.49	&	1.00	&	-53.72	$\pm$	0.25	&	2.90	&	2.9\tablenotemark{c}	&	0.04	&	0.027	\\
G 274-113               	B	&	1.4	&	0.49	&		&	82.68	$\pm$	0.44	&		&		&		&		\\\\
NLTT 6638               	A	&	2.2	&	0.44	&	0.82	&	-23.29	$\pm$	0.67	&	31.10	&	$<$ 356.4	&	$<$0.93	&	$\gg$10	\\
NLTT 6638               	B	&	3.0	&	0.36	&		&	4.34	$\pm$	0.31	&		&		&		&		\\\\
GJ 3129                 	A	&	4.4	&	0.2	&	0.70	&	-37.25	$\pm$	0.47	&	4.00	&	$<$27.8	&	$<$0.13	&	$\gg$10	\\
GJ 3129                 	B	&	5.2	&	0.14	&		&	11.36	$\pm$	0.38	&		&		&		&		\\\\
GJ 3236                 	A	&	3.8	&	0.2	&	1.00	&	45.39	$\pm$	2.90	&	no data	&	$<$20.4	&	$<$0.11	&	$\gg$10	\\
GJ 3236                 	B	&	4.4	&	0.2	&		&	-11.49	$\pm$	2.05	&		&		&		&		\\\\
NLTT 11415              	A	&	1.5	&	0.49	&	0.41	&	11.20	$\pm$	0.17	&	not variable	&	$<$33.2	&	$<$0.18	&	$\gg$10	\\
NLTT 11415              	B	&	3.6	&	0.2	&		&	69.25	$\pm$	0.48	&		&		&		&		\\\\
GJ 3240 Ba              		&	4.2	&	0.2	&	1.00	&	-66.23	$\pm$	2.81	&	not variable	&	$<$0.4	&	0.01	&	1.E-05	\\
GJ 3240 Bb             		&	4.3	&	0.2	&		&	142.72	$\pm$	2.20	&		&		&		&		\\\\
2MASS J04244260-0647313 	A	&	4.5	&	0.17	&	0.77	&	-11.86	$\pm$	0.38	&	no data	&	$<$70.3	&	$<$0.25	&	$\gg$10	\\
2MASS J04244260-0647313 	Ba	&	5.5	&	0.12	&	0.83	&	77.03	$\pm$	0.78	&		&	$<$1.9	&	$<$0.02	&	0.011	\\
2MASS J04244260-0647313 	Bb	&	5.7	&	0.1	&		&	-26.03	$\pm$	0.67	&		&		&		&		\\\\
CCDM J04404+3127 Ba       		&	4.2	&	0.2	&	0.70	&	-7.78	$\pm$	0.64	&	2.048 (EB)	&	2.0	&	0.02	&	0.011	\\
CCDM J04404+3127 Bb      		&	5.0	&	0.14	&		&	91.24	$\pm$	1.00	&		&		&		&		\\\\
GJ 206	A	&	3.9	&	0.2	&	1.00	&	0.93	$\pm$	0.34	&	7.60	&	$<$61.1	&	$<$0.23	&	1.4	\\
GJ 206	B	&	4.3	&	0.2	&		&	40.40	$\pm$	0.52	&		&		&		&		\\\\
GJ 3362	A	&	1.1	&	0.49	&	0.90	&	75.23	$\pm$	1.43	&	no data	&	$<$138.9	&	$<$0.52	&	$\gg$10	\\
GJ 3362	B	&	2.0	&	0.44	&		&	115.01	$\pm$	0.88	&		&		&		&		\\\\
G 108-4                 	A	&	5.0	&	0.14	&	0.71	&	101.78	$\pm$	0.28	&	no data	&	$<$46.9	&	$<$0.16	&	$\gg$10	\\
G 108-4                 	B	&	6.3	&	0.1	&		&	65.41	$\pm$	0.26	&		&		&		&		\\\\
2MASS J07282116+3345127	A	&	3.9	&	0.2	&	1.00	&	-1.30	$\pm$	0.79	&	3.55	&	$<$232.7	&	$<$0.55	&	$\gg$10	\\
2MASS J07282116+3345127	B	&	3.9	&	0.2	&		&	23.98	$\pm$	0.67	&		&		&		&		\\\\
LHS 5134                	A	&	2.4	&	0.44	&	0.82	&	66.52	$\pm$	0.31	&	9.70	&	$<$215.0	&	$<$0.66	&	6.5	\\
LHS 5134                	B	&	3.1	&	0.36	&		&	99.23	$\pm$	0.69	&		&		&		&		\\\\
2MASS J08082487+4347557	A	&	5.1	&	0.14	&	0.71	&	16.21	$\pm$	0.48	&	not variable	&	$<$1.5	&	$<$0.02	&	0.062	\\
2MASS J08082487+4347557	B	&	6.1	&	0.1	&		&	129.70	$\pm$	0.30	&		&		&		&		\\\\
GJ 1108 Ba               		&	2.8	&	0.36	&	1.00	&	-1.06	$\pm$	0.53	&	3.38	&	$<$395.5	&	$<$0.92	&	$\gg$10	\\
GJ 1108 Bb               		&	3.3	&	0.36	&		&	25.30	$\pm$	0.45	&		&		&		&		\\\\
GJ 3547                   	A	&	-0.1	&	0.6	&	1.00	&	27.36	$\pm$	0.93	&	no data	&	$<$20.2	&	$<$0.16	&	$\gg$10	\\
GJ 3547                   	B	&	0.1	&	0.6	&		&	109.68	$\pm$	0.77	&		&		&		&		\\\\
GJ 3630                   	A	&	4.3	&	0.2	&	0.91	&	26.14	$\pm$	1.27	&	not variable	&	$<$6E7	&	$<$2313	&	$\gg$10	\\
GJ 3630                   	Ba	&	5.5	&	0.12	&	0.83	&	-16.58	$\pm$	0.45	&		&	$<$28.1	&	$<$0.11	&	$\gg$10	\\
GJ 3630                   	Bb	&	6.1	&	0.1	&		&	68.06	$\pm$	0.65	&		&		&		&		\\\\
2MASS J12065663+7007514 	A	&	3.9	&	0.2	&	1.00	&	34.13	$\pm$	0.38	&	no data	&	$<$2.7	&	$<$0.03	&	0.024	\\
2MASS J12065663+7007514 	B	&	3.9	&	0.2	&		&	-77.75	$\pm$	0.37	&		&		&		&		\\\\
2MASS J14204953+6049348	A	&	3.2	&	0.36	&	0.56	&	-33.76	$\pm$	0.75	&	no data	&	0.4	&	0.01	&	3E-5	\\
2MASS J14204953+6049348	B	&	3.8	&	0.2	&		&	-12.54	$\pm$	0.39	&		&		&		&		\\\\
GJ 616.2 	A	&	1.1	&	0.49	&	1.00	&	-41.20	$\pm$	0.75	&	not variable	&	$<$530.6	&	$<$1.30	&	$\gg$10	\\
GJ 616.2 	B	&	1.4	&	0.49	&		&	-15.31	$\pm$	0.54	&		&		&		&		\\\\
HAT 199-13890           	A	&	5.0	&	0.14	&	1.00	&	-100.14	$\pm$	0.79	&	not variable	&	$<$0.7	&	$<$0.01	&	0.002	\\
HAT 199-13890           	B	&	5.1	&	0.14	&		&	55.53	$\pm$	0.96	&		&		&		&		\\\\
NLTT 48838              	A	&	3.5	&	0.36	&	0.56	&	-77.56	$\pm$	0.67	&	1.12	&	$<$40.1	&	$<$0.19	&	$\gg$10	\\
NLTT 48838              	B	&	3.8	&	0.2	&		&	-26.75	$\pm$	1.05	&		&		&		&		\\\\
2MASS J21021569-3129118	A	&	4.4	&	0.2	&	1.00	&	-32.70	$\pm$	0.30	&	not variable	&	$<$27.6	&	$<$0.13	&	$\gg$10	\\
2MASS J21021569-3129118	B	&	4.2	&	0.2	&		&	18.74	$\pm$	0.30	&		&		&		&		\\\\
LSPM J2114+1254         	A	&	5.3	&	0.14	&	0.71	&	-104.46	$\pm$	0.31	&	not variable	&	$<$2.9	&	$<$0.03	&	0.792	\\
LSPM J2114+1254         	B	&	5.9	&	0.1	&		&	-12.70	$\pm$	0.40	&		&		&		&		\\\\
BD -225866	Aa\tablenotemark{d}	&	-0.5	&	0.5881	&	1.00	&	-68.97	$\pm$	2.37	&	2.21 (EB)	&	2.21107	&	0.0351	&	0.131	\\
BD -225866	Ab	&	-0.5	&	0.5881	&		&	58.35	$\pm$	0.72	&		&		&		&		\\
BD -225866	Ba	&	1.0	&	0.49	&	0.90	&	-40.23	$\pm$	0.53	&		&	$<$62	&	$<$0.3	&		\\
BD -225866	Bb	&	2.0	&	0.44	&		&	-5.41	$\pm$	0.68	&		&		&		&		\\\\
G 67-46                 	A	&	0.3	&	0.6	&	0.82	&	-31.45	$\pm$	0.78	&	5.00	&	$<$58.9	&	$<$0.31	&	$\gg$10	\\
G 67-46                 	B	&	1.4	&	0.49	&		&	24.39	$\pm$	0.56	&		&		&		&		\\\\
GJ 4359                 	A	&	0.4	&	0.6	&	0.73	&	-59.78	$\pm$	1.02	&	no data	&	$<$3.0	&	$<$0.04	&	0.017	\\
GJ 4359                 	B	&	2.1	&	0.44	&		&	88.89	$\pm$	0.33	&		&		&		&		\\\\
GJ 4362                 	A	&	1.7	&	0.44	&	0.82	&	-10.32	$\pm$	0.14	&	11.15	&	$<$20.8	&	$<$0.14	&	11.403	\\
GJ 4362                 	B	&	3.2	&	0.36	&		&	60.92	$\pm$	0.53	&		&		&		&		\\\\

\enddata

\tablenotetext{a}{Masses were derived from the component spectral types using data from \cite{reid05}.}
\tablenotetext{b}{Rotation periods as observed by WASP photometry. The period uncertainty is $\lesssim$0.1 days.}
\tablenotetext{c}{$P_{max}$ for 2M0008+4757 and G274-113 is 4.63 and 3.63 days, respectively.}
\tablenotetext{d}{Observations of this ESB4 are published in \cite{shko08}.}

\end{deluxetable}

\begin{deluxetable}{lcccccccl}
\tabletypesize{\scriptsize}
\rotate
\tablecaption{Targets with detected rotation period \label{targets_with_Prot}}
\tablewidth{0pt}
\tablehead{
\colhead{Name} & \colhead{N data} &  \colhead{Date begin} & \colhead{Date end} & \colhead{Period\tablenotemark{a}} & \colhead{Amplitude} & \colhead{$\delta\chi^2/\chi^2_{best}$\tablenotemark{a}} & \colhead{Year} & \colhead{Note}    \\
\colhead{} & \colhead{Points} &  \colhead{HJD-2450000} & \colhead{HJD-2450000} &\colhead{days} & \colhead{$\Delta$mag} & \colhead{} & \colhead{} 
}
\startdata


2MASS~J00080642+4757025       &  5381   &  4306.58      & 4451.50  &   4.38   &  0.022      &  0.48  &  2007 &\\  
LHS 6032                      &   4118	&  4332.58      & 4460.52  &   4.05   &  0.008      &  0.36  &  2007  &\\
G 274-113                     &   5856  &  4677.52      & 4813.45  &   2.89   &  0.006      &  0.30  &  2008 &\\
GJ 3129                       &   1009  &  4733.52      & 4767.65  &   4.01   &  0.089      &  1.07  &  2008 &\\
      GJ 206                  &   4662  &  4752.58      & 4863.47  &   7.59   &  0.013      &  0.89  &  2008-9 &\\
2MASS J07282116+3345127        &  9658   &  4056.55      & 4172.50  &   3.57   &  0.028      &  0.69  &  2006  &\\
\phantom{2MASSJ07282116+3345127}&  714   &  4405.58      & 4455.75  &   3.55   &  0.044      &  0.30  &  2007 &\\
\phantom{2MASSJ07282116+3345127}&  602   &  4491.40      & 4554.46  &   3.54   &  0.042      &  1.41  &  2008 &\\
2MASS J08082487+4347557        &   6784  &  4056.58      & 4172.50  &   9.70   &  0.005      &  0.29  &  2006 &\\
GJ 1108 B                     &  5616   &  4056.59      & 4172.50  &   3.37   &  0.030      &  2.48  &  2006 &\\
\phantom{GJ 1108 B}           &   698   &  4056.59      & 4172.50  &   3.37   &  0.048      & 11.46  &  2007 &\\
\phantom{GJ 1108 B}           &   729   &  4491.39      & 4554.42  &   3.40   &  0.027      &  4.38  &  2008 &\\
         NLTT 48838           &  1780   &  4297.45      & 4369.40  &   1.12   &  0.040      &  1.98  &  2007 	&	with no systematics \\
         BD -225866           &  4579   &  3862.58      & 4052.31  &   2.20   &  0.019      &  1.57  &  2006	& Eclipsing \\
\phantom{BD -225866}          &  4557   &  4250.53      & 4418.34  &   2.20   &  0.029      &  4.61  &  2007  &\\
\phantom{BD -225866}          &  5509   &  4622.51      & 4764.41  &   1.10   &  0.006      &  0.21  &  2008	&	 period harmonic detected at 1/2 $P_{orb}$, marginal \\
         G 67-46              &  2555   &  4672.56      & 4767.52  &   5.00   &  0.015      &  1.20  &  2008 &\\
         GJ 4362              &  5068   &  3870.61      & 4054.47  &   11.13  &  0.024      &  1.42  &  2006 &\\
\phantom{GJ 4362}             &  5722   &  4268.53      & 4433.44  &   11.16  &  0.026      &  1.10  &  2007 &\\

\enddata
\tablenotetext{a}{The period uncertainty is $\lesssim$0.1 days.}
\tablenotetext{b}{Here we use the statistics and formalisms of \cite{zechkurst}.  Their figure of merit to determine a variable signal is the improvement in $\chi^2$ with the sinusoidal fit compared to a flat model divided by the $\chi^2$ of the best fitting sinusoidal model.  Real detections have $\delta\chi^2/\chi^2_{best} \geq$ 0.28.}
\end{deluxetable}

\begin{deluxetable}{lcccclll}
\tabletypesize{\scriptsize}
\rotate
\tablecaption{Targets with no detected rotation period \label{targets_without_Prot}}
\tablewidth{0pt}
\tablehead{
\colhead{Name} & \colhead{N data} &  \colhead{Date begin} & \colhead{Date end} & \colhead{$\delta\chi^2/\chi^2_{best}$} & \colhead{Year}  & \colhead{Notes}   \\
\colhead{} & \colhead{Points} &  \colhead{HJD-2450000} & \colhead{HJD-2450000} &\colhead{} & \colhead{} &  \colhead{} & \colhead{} 
}
\startdata
         NLTT 6638            &   11664 &  4677.52      & 4813.45  &  0.13  & 2008 &	VB blended in SWASP data \\
         GJ 3240              &   2269  &  3995.58      & 4125.39  &  0.13  & 2007 & blended with J033733.35+175114.3 \\
         NLTT 11415           &   2269  &  3995.58      & 4125.39  &  0.12  & 2007 & blended with J033733.88+175100.6 \\
2MASS J04244260-0647313       &   588   &  4491.32      & 4503.46  &  0.14  & 2008, early &\\
\phantom{2MASSJ04244260-0647313}& 692   &  4721.50      & 4760.75  &  0.01  & 2008, late &\\
CCDM J04404+3127B             &   949   &  3219.73      & 3278.76  &  0.10  & 2004 &	Eclipsing, no detectable rotational signal, blended  \\
\phantom{CCDM J04404+3127B}   &   2524  &  3995.59      & 4147.48  &  0.06  & 2006 &\\  
\phantom{CCDM J04404+3127B}   &   1189  &  4372.59      & 4460.64  &  0.06  & 2007 &\\ 
2MASS J08082487+4347557       &   1490  &  4419.60      & 4455.75  &  0.02  & 2007 &\\
\phantom{2MASSJ08082487+4347557}& 3865  &  4491.39      & 4554.43  &  0.02  & 2008 &\\
         GJ 3630              &   2071  &  4501.52      & 4581.44  &  0.03  & 2008 &\\
         GJ 616.2             &   536   &  3920.38      & 3950.52  &  0.09  & 2006 & \\
\phantom{GJ 616.2}            &   10581  & 4189.59      & 4316.50  &  0.37  & 2007 & 1.99d period, systematics present \\
\phantom{GJ 616.2}            &   8051   & 4553.61      & 4681.52  &  0.17  & 2008 &\\
         HAT 199-13890        &   1686   & 4230.59      & 4296.39  &  0.04  & 2007 &\\
2MASS J21021569-3129118       &  11471   & 3862.55      & 4037.38  &  0.03  & 2006 &\\
\phantom{2MASS J21021569-3129118}& 1221  & 4236.50      & 4399.41  &  0.10  & 2007 & 1 day period  \\
CCDM J2114+1254               &   8715   & 3128.66      & 3278.56  &  0.04  & 2004  &  \\
\phantom{CCDM J2114+1254}     &   7953   & 3908.55      & 4023.48  &  0.03  & 2006  &\\

\enddata
\end{deluxetable}


 \clearpage

\end{document}